\documentclass{ws-ijqi}
\newtheorem{inference}{Inference}

\begin{document}
\markboth{R. Srinivasan}{Bohr's Complementarity Principle Under The NAFL Interpretation}
\title{LOGICAL ANALYSIS OF THE BOHR COMPLEMENTARITY PRINCIPLE IN AFSHAR'S EXPERIMENT UNDER
THE \\NAFL INTERPRETATION}

\author{RADHAKRISHNAN SRINIVASAN}

\address{IBM India Software Labs, Embassy Tech Zone, Building ETB1-5F-A, Phase 2, Rajiv Gandhi Infotech Park, Hinjewadi,
Pune 411057, India \\ sradhakr@in.ibm.com}

\maketitle

\begin{history}
\received{11 July 2005}
\revised{6 November 2008 and 26 January 2010}
\end{history}

\begin{abstract}
The NAFL (non-Aristotelian finitary logic) interpretation of quantum mechanics
requires that no `physical' reality can be ascribed to the wave nature of the
photon. The NAFL theory QM, formalizing quantum mechanics, treats the superposed state~($S$)
of a single photon taking two or more different paths at the same time as a logical
contradiction that is formally unprovable in QM. Nevertheless, in a nonclassical NAFL
model for QM in which the law of noncontradiction fails, $S$ has a meaningful
metamathematical interpretation that the classical path information for the photon is not available.
It is argued that  the existence of an interference pattern does not logically amount to a proof of the self-interference of
a single photon. This fact, when coupled with the temporal nature of NAFL truth, implies the logical
validity of the retroactive assertion of the path information (and the logical superfluousness of the grid) in Afshar's experiment.
The Bohr complementarity principle, when properly interpreted with the time dependence of logical truth taken into account, holds in Afshar's experiment.
NAFL supports, but not demands, a metalogical reality for the particle nature of the photon even when the semantics of QM requires the state~$S$.
\end{abstract}


\section{Introduction}\label{intro}
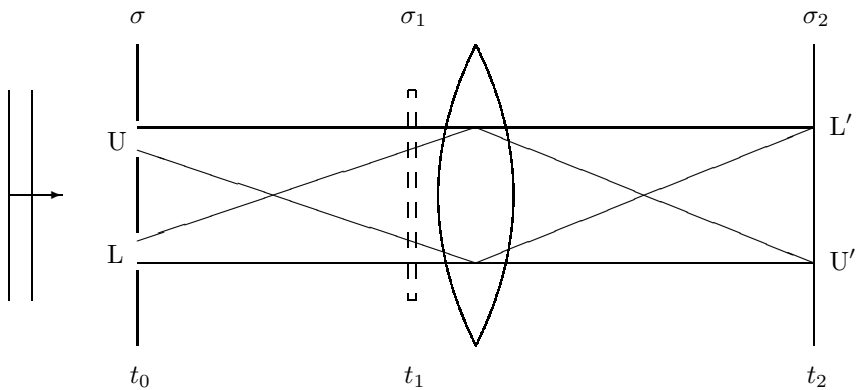
\begin{figure}\label{f1}
\setlength{\unitlength}{1cm}
\begin{picture}(11,5)
\put(1.7,0.5){\line(0,1){1}}
\put(1.7,2){\line(0,1){1}}
\put(1.7,3.5){\line(0,1){1}}
\put(10.7,0.5){\line(0,1){4}}
\qbezier(6.2,0.5)(5.2,2.5)(6.2,4.5)
\qbezier(6.2,0.5)(7.2,2.5)(6.2,4.5)
\put(10.9,1.5){$\mbox{U}^{\prime}$}
\put(10.9,3.3){$\mbox{L}^{\prime}$}
\put(1.7,1.6){\line(1,0){9}}
\put(1.7,3.4){\line(1,0){9}}
\put(6.2,1.6){\line(5,2){4.5}}
\put(6.2,1.6){\line(-3,1){4.5}}
\put(6.2,3.4){\line(5,-2){4.5}}
\put(6.2,3.4){\line(-3,-1){4.5}}
\put(0,1.1){\line(0,1){2.8}}
\put(0.3,1.1){\line(0,1){2.8}}
\put(0,2.5){\vector(1,0){0.7}}
\put(5.3,1.1){\dashbox{0.2}(0.1,2.8)}
\put(1.3,1.6){L}
\put(1.3,3.1){U}
\put(1.6,4.8){$\sigma$}
\put(5.2,4.8){$\sigma_{1}$}
\put(10.55,4.8){$\sigma_{2}$}
\put(1.6,0){$t_{0}$}
\put(5.25,0){$t_{1}$}
\put(10.6,0){$t_{2}$}
\end{picture}
\caption{Sketch of Afshar's experiment}
\end{figure}

Shahriar~S.~Afshar~\cite{afshar1,afshar2,afshar3} has recently performed a variant of
Young's two-slit experiment, in which he claims to have demonstrated the
falsification of the celebrated Bohr complementarity principle, and thereby, the Copenhagen
interpretation of quantum mechanics. In Afshar's experiment, sketched in Fig.~\ref{f1},
a source emits photons towards a screen with two slits marked U and L.
A suitably located converging lens downstream of the screen focuses the photon streams
into sharp images of the two slits U and L, marked  $\mbox{U}^{\prime}$ and
$\mbox{L}^{\prime}$ respectively, at location $\sigma_{2}$ (for simplicity,
we do not distinguish between the image plane and the focal plane in Fig.~\ref{f1}). Afshar notes
that these images serve to provide sharp which-way information for the photons, thereby
confirming their particle status. A wire grid is placed at location $\sigma_{1}$
in Fig.~\ref{f1}, such that the (thin) wires occur at precisely the theoretically
predicted minima of the interference pattern due to the superposed
states of the photons. Afshar observes that the wire grid effectively performs
a nondestructive confirmation of the superposition; when both slits are open,
no distortion or reduction in intensity of the images at $\sigma_2$ is observed
as compared to the case when the wire grid is removed. The expected distortion and
reduction in intensity do occur when one of the slits is closed. This seemingly rules out
the classical particle state of the photons between the locations~$\sigma$ and $\sigma_{1}$
when both slits are open, and confirms their wave nature, since blocking of photons by the
wire grid does not occur to the classically expected extent. Afshar concludes that both
the sharp particle and wave natures of the photon are exhibited in a single experiment,
in violation of the Bohr complementarity principle. In particular, Afshar~\cite{afshar1} claims that
the Englert-Greenberger duality relation~\cite{greenberger,englert} is falsified in his experiment.

Afshar's interpretation of his experiment has many critics (see Refs.~\refcite{kastner1}--\refcite{steuernagel}),
but there is no consensus on why Afshar is supposedly wrong. The most prominent of these critics seems to be
Kastner~\cite{kastner1}, who rejects the existence of which-way information in not only the Afshar experiment,
but also in any classical two-slit experiment even when the wire grid is not present. Kastner's argument is
essentially based on a realist interpretation in which the presence of an interference pattern requires the photon
to `really' pass through both slits. Kastner concludes that post-selection of the photon in a `which-slit' basis does
not imply that the photon passed through only one slit. In a subsequent article, Kastner~\cite{kastner2} cites the two-slit
experiment of Srikanth~\cite{srikanth} as even more dramatic than the Afshar experiment, in that the existence of the
interference pattern is irreversibly recorded (rather than just inferred) and yet post-selection of the photon in a `which-slit'
basis is possible~\footnote{An anonymous referee has rejected the validity of Srikanth's experiment}. The following
passage from Kastner~\cite{kastner2} clearly reveals her realist argument:
\begin{quote}
This is even more dramatic than the Afshar result because clearly $V=1$ since a fully articulated interference pattern
has  been irreversibly recorded -- not just indicated indirectly -- and yet a measurement can be performed after the fact
that seems to  reveal `which slit' the photon went through. However, the point is that the detector's vibrational
mode remains in a  superposition until that measurement is made, implying that each photon indeed went through both
slits. As Srikanth puts  it,`...the amplitude contributions from both paths to the observation at [detector element]
$x$ results in a superposition of vibrational modes. The initial superposition leaves behind a remnant
superposition.'~\cite{srikanth}. So, just because one can `post-select' by  measuring the vibrational observable and
end up with a particular corresponding slit eigenstate doesn't mean the particle  actually went through that slit;
in a very concrete sense, it went through both slits.
\end{quote}
Kastner~\cite{kastner1,kastner2} concludes that the Bohr complementarity principle is not violated in the Afshar
experiment and that the Englert-Greenberger duality relation is not applicable in all similar experiments in which an interference pattern
can be inferred without `collapsing' the superposed state. Other critics~\cite{qureshi,reitzner} of the Afshar experiment agree with Kastner
that there is no which-way information and like Kastner, argue for a realist interpretation in which the photon
passed through both slits. However, a few workers~\cite{drezet,steuernagel} assert that while which-way information is
present in the Afshar experiment, visibility of the fringes is very small because the interference pattern is
not actually recorded in the Afshar experiment. This group rejects the inference of the existence of an interference
pattern and insists that only an experimental recording of the same is acceptable. Afshar's co-workers~\cite{flores}
and O'Hara~\cite{ohara} support the Afshar interpretation that both the particle and wave properties of the photon
co-exist in a real sense, thereby rejecting both the Bohr complementarity principle and the Englert-Greenberger
duality relation. 

It is the purpose of this paper to analyze the status of the Bohr complementarity
principle in Afshar's experiment, particularly in the light of the non-Aristotelian
finitary logic~(NAFL) recently proposed by the author~\cite{ijqi,1166,acs,ract}.
We will not concern ourselves with the nuts and bolts of Afshar's experiment
in an attempt to evaluate his claims from the point of view of experimental physics.
Instead, we consider the experiment from a purely logical angle, in the case
when Afshar's claims are granted as experimentally sound. It will be seen that
the NAFL interpretation of the quantum superposition principle does indeed uphold
complementarity, \emph{despite} the co-existence of the interference pattern and the path information for the photons that arrive at the image plane.
Indeed, we argue that the interference pattern should, in principle, be reconstructable for these photons from just the path information, without any need
for the controversial grid in Afshar's experiment. In fact the grid is not only logically superfluous, but also does not provide a complete
reconstruction of the interference pattern, as would be possible from analysis of the path information.
Kastner's analysis~\cite{kastner1,kastner2} is critically examined and her claim that Cramer's Transactional
Interpretation of quantum mechanics~\cite{cramer} rescues the complementarity principle in Afshar's
experiment is disputed. What the experiments of Afshar and others
really establish is that any realist interpretation of quantum mechanics is highly problematic from a
physical and philosophical point of view. Both the Bohr complementarity principle
and the Englert-Greenberger duality relation (at its extreme ends) are upheld in the Afshar experiment
when they are properly interpreted to take into account the time dependence of logical truth, as embodied by NAFL. The NAFL
interpretation really vindicates the original non-realist Copenhagen interpretation championed by
Neils Bohr and others, which, contrary to Afshar's claims, remains unscathed by the Afshar experiment.
While the Copenhagen interpretation formulates Bohr complementarity as a physical principle, the NAFL
interpretation enshrines it as a sacred and inviolable logical principle which follows from basic
postulates that embody finitary reasoning.

\section{Summary of the main argument and conclusions}\label{summary}
This section highlights the main argument and conclusions of this paper; the details are presented in subsequent
sections. Consider the modified experiment of Afshar~\cite{afshar3}, which was originally proposed
by Wheeler~\cite{wheeler} and is described in Sec.~5 of Ref.~\refcite{flores}, as quoted below.
\begin{quote}
\dots~a laser beam impinges on a 50:50 beam splitter and produces two spatially separated coherent  beams of equal
intensity. The beams overlap at some distance, where they form an interference  pattern of bright and dark fringes.
At the center of the dark fringes we place thin  wires.~\dots~Beyond the region of overlap the two beams fully
separate again. There, two detectors are positioned such  that detector~$1^{\prime}$ detects only the photons originating
from mirror~1, and detector~$2^{\prime}$ detects only photons originating from the beam splitter (mirror~2).
Since the pathway of the photon is practically unobstructed, a study of the electric  fields involved together with
conservation of momentum allows us to uniquely identify, with high  probability, the respective mirror as the place
where that photon originated.
\end{quote}
The wire grid in the modified experiment plays the same role as in the original Afshar experiment and
Afshar~\cite{afshar3} concludes that the results of the two experiments are in agreement.

The above experiment is similar in essence to the quantum eraser, the delayed-choice version of which
is particularly interesting~\cite{scully}. When the two photon beams overlap, they contain interference information
and when they separate, they contain which-way information. As long as no measurements are made, the available
information may be reversed as many times as one chooses. The key point here is that \emph{when} the photon beams provide
interference information, there is no way to extract which-way information and vice versa; this is essentially
a formulation of the Bohr complementarity principle and is fully in agreement with the Copenhagen
interpretation.~\footnote{The author is grateful to an anonymous referee for suggesting this
formulation and the analogy of the quantum eraser for the Afshar experiment.}
The emphasis on `when' in this formulation is particularly important and contains within it a tacit
time-dependence that has not been explored in full depth. The only new element in the Afshar experiment is
the presence of the wire grid. As in Fig.~\ref{f1},  let the photon beams converge at a spatial location $\sigma_1$, where they
pass through the wire grid. At location $\sigma_2$ let the photon beams hit the detectors, after which
one is able to deduce both the existence of the interference pattern and the path information
for the photons.  We argue below that these facts do not constitute a violation of the Bohr complementarity principle as claimed
by Afshar~\cite{afshar1,afshar2,afshar3}. Further, the wire grid is superfluous in the sense that it is not required
to deduce the existence of the interference pattern; the complete path information is sufficient for this purpose.

Consider a photon that passes through the wire grid at time~$t_1$ and then reaches a detector at time~$t_2$. Afshar's argument is essentially that
a large number of such photons that reach the detectors at time~$t_2$ must have formed an interference pattern at time $t_1$, because of the negligibly small
number of photons that impinged on the wire grid at time~$t_1$ (as compared to the number that reached the detectors).  It is crucial to note that Afshar is able to argue
for the existence of the interference pattern for this large number of photons only \emph{after} they have reached the detectors, at which point of time complete path
information is also available for each of these photons. If these photon paths are superposed on to an imaginary screen at location~$\sigma_1$, a complete reconstruction
of the interference pattern is theoretically possible, without any need for the wire grid. Therefore if one accepts Afshar's argument that the Bohr complementarity principle
is violated in his experiment, one must also accept the same conclusion from a classical version of his experiment, in which the wire grid is not present. In fact the analysis
of the path information would provide a complete reconstruction of the interference pattern, whose existence is only nonconstructively deduced via the wire grid. Hence
the wire grid is not only logically superfluous in Afshar's argument, but it also provides a less satisfactory and controversial method for deducing the existence of an interference pattern.
It is extremely important to note that Afshar also accepts the validity of the path information for a photon in his experiment, post its detection at time~$t_2$. It follows that Afshar has also
essentially analyzed particle-like photons, each of which he accepts as having passed through only one of the slits, to arrive at his
conclusion that an interference pattern existed at a previous time~$t_1$ for these photons.

We may conclude that Afshar's method of deducing (or as he claims, measuring) the existence of an interference pattern does not provide any evidence whatsoever
of wave-like properties for a photon. Likewise, analysis of the path information in a classical version of Afshar's experiment (without the wire grid)
would also provide a reconstruction of the interference pattern without providing any evidence whatsoever of wave-like properties for a photon.
In contrast, the classical method for measuring an interference pattern, via a destruction of the photons by a screen located at $\sigma_1$, would eliminate
any possibility of deducing path information for these photons. It is the non-availability of the path information at time $t_1$ that establishes the wave nature of the photon
(via the superposed state $S$), whose logical consequence is the presence of an interference pattern. It is extremely important to note that the converse implication does not hold:
the presence of an interference pattern does not logically imply that that the photons are in a superposed state. Hence the co-existence of the interference pattern and the path
information, as deduced at time $t_2$, is not a contradiction as claimed by Afshar. Post the time $t_2$, the correct interpretation of the Afshar experiment (or a classical version
without the wire grid) is that particle-like photons have a non-classical probability distribution, resulting in an interference pattern. Here one must understand the time dependence
involved and give up any realist notions of the superposed state $S$, or the wave nature of the photon.

The state $S$ should be thought of as merely a formalism by which we may deduce probability distributions, which
are confirmed by the existence of the interference pattern. The Copenhagen interpretation championed by Bohr has
always been non-realist in this sense. As we will see in the ensuing sections, in the (non-realist) NAFL
interpretation, the state $S$ of the photons has the precise logical meaning that path information is not available;
it does \emph{not} mean that a single photon `really' took both paths, which is a logical impossibility for a particle.
Secondly, the retroactive assertion of the path information at time~$t_2$ is a logical truth that \emph{only}
applies for times $t \ge t_2$; it does \emph{not} mean that the path information was always available. Again, one needs
a non-classical, temporal logic like NAFL to formulate this time-dependence, which, however, was always tacitly
implied by the Copenhagen interpretation. Seen in this light, we know the following facts about the modified
Afshar experiment. Firstly, at time $t_1$, path information is not available for the photons that are passing
through the wire grid (and which will impinge on the detectors at time $t_2$), and hence they are in the superposed state
$S$ at that instant ($t_1$). The interference pattern is a logical consequence of this state. Secondly, at time $t_2$,
path information is retroactively available for the photons that impinged on the detectors at that instant;
however, this retroactive assertion itself only applies for times $t \ge t_2$. The inferred
presence of the interference pattern does not logically imply that the photons are in a superposed
state $S$; it is this fact which permits a retroactive assertion of the path information post the time $t_2$.

A contradiction would ensue from the above two facts if one insists on ascribing reality for the
superposed state $S$ and also insists that the retroactive assertion of the path information was always
applicable (as would be the case in classical logic). In the first case one would have the contradiction that
the photons `really' took both paths \emph{and} the photons took only one of the available paths; in the second case the
contradiction would be that path information is both available and not available for the said photons. Kastner's
realist stance~\cite{kastner1,kastner2} leads to these contradictions and hence forces her to reject
the retroactive assertion of the path information even in the case when the wire grid is not present (as noted in Sec.~\ref{intro}).
This stance contradicts well-accepted results in physics and fails to explain how a photon can `really' take both paths and yet exhibit
particle-like behaviour at the detectors (decoherence).

The Bohr complementarity principle is formulated in the NAFL interpretation as a logical principle (see
Sec.~\ref{afsh}). Namely, that at any given time, a photon can either be in a superposed state, in which
case the law of noncontradiction fails, or in a classical particle-like state, when the law of noncontradiction
applies; at any given time, it is not logically possible for the photon to be in both of these states. This
formulation reduces to that noted earlier in terms of which-way information and interference information.
The Englert-Greenberger duality relation~\cite{greenberger,englert}, when properly interpreted, is also
upheld in the Afshar experiment, contrary to Afshar's claims~\cite{afshar1,afshar2,afshar3}. This relation
may be expressed in the form
\begin{equation}
D^2 + V^2 \le 1. \label{eg}
\end{equation}
Here $V$ is the visibility of the interference fringes and $D$ is the distinguishability of the photon paths.
Afshar claims that in his experiment, both $D$ and $V$ are close to 1, and hence Eq.~(\ref{eg}) is violated.
However, if the time dependence of the parameters $D$ and $V$ is taken into account,
one may conclude that Eq.~(\ref{eg}) is indeed upheld in the Afshar experiment. At time~$t_1$, when the photons
pass through the wire grid, there is no path information and hence $D=0$ for these photons; further, the photons are in
a superposed state at this time with the interference fringes present, and hence $V=1$. Indeed, this is certainly true for the
photons that impinge on the wire grid. Post the time $t_2$, path information is available for the photons that impinge
on the detectors, and hence we take $D=1$ for these photons. As noted
earlier, we may infer that these photons were part of an interference pattern at time $t_1$, regardless of whether
the wire grid is present. Despite this inference, one concludes that $V=0$ for time $t \ge t_2$, because the presence of the
interference pattern does not imply that the photons are in a superposed state and so the
path information stands for these times. And when path information is available, the interference pattern is no longer a logical
consequence of any wave-like properties of the photon, which is why we take $V=0$ at these times. When $D$ and $V$ are interpreted in this manner, it is impossible for
Eq.~(\ref{eg}) to be violated in the Afshar experiment (with or without the wire grid) because it expresses precisely the Bohr
complementarity principle.

Finally, consider the analogy of the Schr\"odinger cat experiment (see Sec.~\ref{sch}). At time $t_1$,
let us say that the cat is in the box and is in a superposed state of `alive and dead'. What this state means
in the NAFL interpretation is that no information is available at time~$t_1$ as to the cat's classical state
(`alive' or `dead'). Subsequently, at time~$t_2$, when the box is opened, suppose the cat is found in the `alive'
state. One may retroactively infer at time~$t_2$ that the cat was alive at time~$t_1$. This retroactive inference
only applies for times~$t \ge t_2$ and does not contradict the fact that at time~$t_1$, the cat was in a superposed
state (to which no `reality' can be ascribed). Again the complementarity principle is upheld in both the 
NAFL and the Copenhagen interpretations because at any given time, the cat is in only one state, despite the
retroactive inference. Whereas Kastner's interpretation~\cite{kastner1,kastner2} would
amount to barring the retroactive inference that the cat was in the `alive' state at time~$t_1$ because it `really'
was in the superposed state of `alive and dead' at that time. Any such `reality' is obviously aphysical and
inexplicable, and therefore untenable in our view.

\section{The NAFL interpretation of quantum superposition}\label{qsnafl}
At the outset, we hasten to note that the NAFL interpretation is still nascent and
incomplete in the sense that a lot of work remains to be done in demonstrating how
real analysis can be done in NAFL~\cite{ract}, and ultimately, how all of quantum mechanics can
be formalized in this logic. What has been accomplished at this stage is a completely
new and logical interpretation of some of the ``weird'' phenomena of quantum mechanics,
in particular, superposition and entanglement~\cite{ijqi,1166,acs}. In this section, we will
confine ourselves to a brief exposition of the NAFL interpretation of quantum superposition,
and refer the reader to the original references for further details. The reader who
is already familiar with these details may skip to Sec.~\ref{bcnafl}. Our purpose herein
is to provide just enough information on NAFL so as to enable an appreciation
of the delicate and subtle logical issues involved in the interpretation of the Bohr
complementarity principle in Afshar's experiment, which is discussed in Sec.~\ref{bcnafl}.

The language, well-formed formulae and rules of inference of
NAFL theories~\cite{ijqi} are formulated in exactly the same manner as in classical
first-order predicate logic with equality~(FOPL), where we shall assume, for convenience,
that natural deduction is used; however, there are key differences and restrictions
imposed by the requirements of the Main Postulate of NAFL, which is explained in this
section. In NAFL, truths for \emph{formal propositions} can exist \emph{only} with respect
to axiomatic theories. There are no absolute truths in just the \emph{language} of
a NAFL theory, unlike classical/intuitionistic/constructive logics. There do exist
absolute (metamathematical, Platonic) truths in NAFL, but these are truths
\emph{about} axiomatic theories and their models. As in FOPL, a NAFL theory T
is defined to be consistent if and only if T has a model, and a proposition
$P$ is undecidable in T if and only if neither $P$ nor its negation
$\neg P$ is provable in T. 

\subsection{The Main Postulate of NAFL}\label{mp}
If a proposition $P$ is provable/refutable in a consistent NAFL theory T, then $P$
is true/false with respect to T (henceforth
abbreviated as `true/false in T'); \emph{i.e.}, a model for T will assign $P$ to be
true/false. If $P$ is undecidable in a consistent NAFL theory T, then the Main
Postulate~\cite{1166} provides the appropriate truth definition as follows: $P$ is true/false
in T if and only if $P$ is provable/refutable in an \emph{interpretation} T* of T. Here T*
is an axiomatic NAFL theory that, like T, temporarily resides in the human mind and acts as a
`truth-maker' for (a model of) T. The theorems of T* are precisely those propositions
that are assigned `true' in the NAFL model of T, which, unlike its classical counterpart,
is not `pre-existing' and is instantaneously generated by T*. It follows that T* must necessarily
prove all the theorems of T. Note that for a given consistent theory T, T* could vary in time according to the
free will of the human mind that interprets T; for example, T* could be T+$P$ or
T+$\neg P$ or just T itself at different times for a given human mind, or in the context
of quantum mechanics, for a given �observer�. Further, T* could vary from one observer
to another at any given time; each observer determines T* by his or her own free will.
The essence of the Main Postulate is that $P$ is true/false in T if and only if it has been
\emph{axiomatically declared} as true/false by virtue of its provability/refutability
in T*. In the absence of any such axiomatic declarations, \emph{i.e.}, if $P$ is
undecidable in T* (\emph{e.g.}~take T*=T), then $P$ is `neither true nor false' in T and
Proposition~\ref{p1} shows that consistency of T requires the laws of the excluded middle and
noncontradiction to fail in a nonclassical model for T in which $P \& \neg P$ is the case.
\begin{proposition}\label{p1}
Let $P$ be undecidable in a consistent \emph{NAFL} theory \emph{T}. Then $P \vee \neg P$
and $\neg (P \& \neg P)$ are not theorems of \emph{T}. There must exist a nonclassical
model $\mathcal{M}$ for \emph{T} in which $P \& \neg P$ is the case.
\end{proposition}
For a proof of Proposition~\ref{p1}, see Ref.~\refcite{ijqi} or Appendix~A of Ref.~\refcite{acs};
this proof also seriously questions the logical/philosophical basis for the law
of noncontradiction in both classical and intuitionistic logics.
The interpretation of $P \& \neg P$ in the nonclassical model will be explained
in Sec.~\ref{qs}. Proposition~\ref{p1} is a metatheorem, \emph{i.e.}, it is a theorem
\emph{about} axiomatic theories. The concepts in Proposition~\ref{p1}, namely, consistency,
undecidability (or provability) and the existence of a nonclassical model for a theory
and hence, quantum superposition and entanglement, are strictly metamathematical
(\emph{i.e.}, pertaining to semantics or model theory) and not formalizable in the
syntax of NAFL theories. A NAFL theory~T is either consistent or inconsistent,
and a proposition~$P$ is either provable or refutable or undecidable in T, \emph{i.e.},
the law of the excluded middle applies to these metamathematical truths.
Note that the existence of a nonclassical model does not make T inconsistent or even
paraconsistent in the conventional sense, because T does not
\emph{prove} $P \& \neg P$. However, one could assert that the model theory for T requires the framework
of a paraconsistent logic, so that the nonclassical models can be analyzed. NAFL is the only logic that
correctly embodies the philosophy of formalism~\cite{1166}; NAFL truths for formal propositions are
axiomatic, mental constructs with strictly no Platonic world required.

\subsection{Quantum superposition justified in NAFL}\label{qs}
The nonclassical model $\mathcal{M}$ of Proposition~\ref{p1} is a superposition of two or
more classical models for T, in at least one of which $P$ is true and $\neg P$ in another.
Here `(non-\nolinebreak)classical' is used strictly with respect to the status of $P$. In
$\mathcal{M}$, `$P$'~(`$\neg P$') denotes that `$\neg P$'~(`$P$') is not provable in T*, or
in other words, $\mathcal{M}$ expresses that neither $P$ nor $\neg P$ has been
axiomatically declared as (classically) true with respect to T; thus $P$, $\neg P$,
and hence $P \& \neg P$, are indeed (nonclassically) true \emph{in our world}, according
to their interpretation in $\mathcal{M}$. Note also that $P$ and $\neg P$ are
\emph{classically} `neither true nor false' in $\mathcal{M}$, where `true' and `false' have
the meanings given in the Main Postulate. The quantum superposition principle is justified
by identifying `axiomatic declarations' of truth/falsity of $P$ in T (via its
provability/refutability in T* as defined in the Main Postulate) with `measurement' in the
real world. NAFL is more in tune with the Copenhagen interpretation of quantum mechanics
than the many-worlds interpretation~(MWI). Nevertheless, the \emph{information content} in
$\mathcal{M}$ is that of two or more classical models (or `worlds'), and MWI is at least
partially vindicated in this sense.

\subsection{Example: Schr\"odinger's cat}\label{sch}
Consider the situation wherein the cat is put into the box at time $t=t_0$ and has a
probability $0.5$ of being in the `alive' state at $t=t_2$, when a `measurement'
is made of its state. Let $P$ be the proposition that `The cat is alive', with $\neg P$
denoting `The cat is dead'; obviously, $P$ is undecidable in a suitable formalization QM
of quantum mechanics, which may be taken to include definitions describing this experiment. For
$t_0 < t < t_2$, the observer makes no measurements, and in tune with the
identification noted in Sec.~\ref{qs}, makes no axiomatic declarations regarding $P$
in the interpretation QM* (say, let QM*=QM for this time period). In the resulting
nonclassical model $\mathcal{M}$ of QM, the superposed state $P \& \neg P$ is the case;
this means that the cat has not been declared (measured) to be either alive or dead, which
is certainly true in the real world. At $t=t_2$, if $P$~($\neg P$) is observed,
then the observer takes, say, QM*=QM+$V$~($\neg V$), where $V$~($\neg V$) is defined as
`The cat is alive~(dead) at $t=t_2$'; note that QM* will prove
$P$~(or $\neg P$) at $t=t_2$, \emph{i.e.}, when the observer measures the cat to be
alive~(dead) in the real world, he makes the appropriate axiomatic declarations in his
mind, thus setting up QM* as defined. It should be emphasized that a NAFL theory only
`sees' the observer's axiomatic declarations and does not care whether the real world
exists. The observer sees the real world and the proposed identification of his
measurements with his axiomatic declarations is only an informal convention that is outside
the purview of NAFL. The observer could also use his free will to make his axiomatic
declarations irrespective of (and possibly in contradiction to) what he measures in the
real world; of course, if $P$ is not about the real world, then he has no other choice.
NAFL correctly handles the temporal nature of truth via the time-dependence of QM*.
If $P$ is observed (and axiomatically asserted via QM*=QM+$V$) at $t=t_2$, then the
proposition $U$ that ``The cat was alive for $t_0 < t < t_2$'' can be formalized for
$t \ge t_2$ and proven in the NAFL theory QM*=QM+$V$; $U$ does not conflict temporally
or in meaning with the superposed state $P \& \neg P$, which applies for $t_0 < t < t_2$.
We will return to the Schr\"odinger cat example in Sec.~\ref{ctoss}, in order to further
elaborate upon the validity of this retroactive assertion of $U$.

\subsection{Theory syntax and proof syntax}\label{tsps}
A NAFL theory T requires two levels of syntax, namely the `theory syntax' and the
`proof syntax'. The theory syntax consists of precisely those propositions that are
legitimate, \emph{i.e.}, whose truth in T satisfies the Main Postulate; obviously,
the axioms and theorems of T are required to be in the theory syntax. Further, one can
only add as axioms to T those propositions that are in its theory syntax. In particular,
neither $P \& \neg P$ nor its negation $P \vee \neg P$ is in the theory syntax when
$P$ is undecidable in T. The proof syntax, however, is classical because NAFL has the
same rules of inference as FOPL; thus $\neg (P \& \neg P)$
is a valid deduction in the proof syntax and may be used to prove theorems of T. For
example, if one is able to deduce $A \Rightarrow P \& \neg P$ in the proof syntax of T
where $P$ is undecidable in T and $A$ is in the theory syntax, then one has proved $\neg A$
in T despite the fact that $\neg (P \& \neg P)$ is not a theorem (in fact not even a
legitimate proposition) of T. This is justified as follows: $\neg (P \& \neg P)$ may be
needed to prove theorems of T, but it does not follow in NAFL that
the theorems of T imply $\neg (P \& \neg P)$ if $P$ is undecidable in T.
Let $A$ and $B$ be undecidable propositions in the theory syntax of T.
Then $A \Rightarrow B$ (equivalently, $\neg A \vee B$) is
in the theory syntax of T if and only if $A \Rightarrow B$ is \emph{not}
(classically) deducible in the proof syntax of T. It is easy to check that if
$A \Rightarrow B$ is deducible in the proof syntax of T, then its (illegal) presence in the
theory syntax would force it to be a theorem of T, which is not permitted by
the Main Postulate. For, in a nonclassical model $\mathcal{M}$ for T in which both $A$
and $B$ are in the superposed state, $A \& \neg B$ must be nonclassically true, but
theoremhood of $A \Rightarrow B$ will prevent the existence of $\mathcal{M}$. If one
replaces $B$ by $A$ in this result, one obtains the previous conclusion that
$\neg (A \& \neg A)$ is not in the theory syntax. For example, take $\mbox{T}_0$ to be the
null set of axioms. Then nothing is provable in $\mbox{T}_0$, \emph{i.e.}, every legitimate
proposition of $\mbox{T}_0$ is undecidable in $\mbox{T}_0$. In particular, the proposition
$(A \& (A \Rightarrow B))\Rightarrow B$, which is deducible in the proof syntax of $\mbox{T}_0$
(via the \emph{modus ponens} inference rule), is not in the theory syntax; however,
if $A \Rightarrow B$ is not deducible in the proof syntax  of $\mbox{T}_0$,
then it is in the theory syntax. Note also that $\neg \neg A \Leftrightarrow A$ is
not in the theory syntax of $\mbox{T}_0$; nevertheless, the `equivalence' between
$\neg \neg A$ and $A$ holds in the sense that one can be replaced by the other
in every model of $\mbox{T}_0$, and hence in all NAFL theories. Indeed, in a nonclassical
model for $\mbox{T}_0$, this equivalence holds in a nonclassical sense and must be
expressed by a different notation~\cite{1166}.

\section{Bohr Complementarity and the NAFL interpretation of Afshar's experiment}\label{bcnafl}
The Bohr complementarity principle easily follows from the NAFL interpretation of
quantum superposition discussed in Sec.~\ref{qsnafl}. As applied to Afshar's experiment,
the relevant definition is as follows.
\begin{definition}[Bohr complementarity principle~(BCP)]\label{bcp}
The particle and wave nature of the photon cannot be \emph{simultaneously}
demonstrated to hold at any given spatial location and time within a given experiment.
\end{definition}
Note carefully the location of the word ``\emph{simultaneously}'' in this definition;
if we were to change BCP to ``\dots cannot be demonstrated to hold \emph{simultaneously}
at any given spatial location and time within a given experiment'', then such a
definition would arguably fail even in NAFL. The ability of NAFL to handle the
temporal nature of mathematical truth and the distinctions NAFL makes between
syntax and semantics are demonstrated to be very important for a
correct logical explanation of the results in Afshar's experiment.

\subsection{Quantum superposition in Afshar's experiment}\label{qsae}
Let QM be the NAFL theory formalizing quantum mechanics. We assume that definitions providing
a detailed description of the single-photon version of Afshar's experiment~\cite{afshar1,afshar2,afshar3}
(which he has reportedly performed with the same results) have already been included in QM.
\begin{definition}\label{defp}
Let $P$ denote the proposition ``The photon passed through (only) slit~U
at location~$\sigma$ and time~$t_0$ in Fig.~\ref{f1}.''
\end{definition}
\begin{definition}\label{defnegp}
Let the negation $\neg P$ of $P$ denote the proposition ``The photon passed through
(only) slit~L at location~$\sigma$ and time~$t_0$ in Fig.~\ref{f1}.''
\end{definition}
Since $P$ and $\neg P$ are equally probable, it follows that $P$ is undecidable in QM.
Here it is important to understand the NAFL concept of negation (see Sec.~2.2 of
Ref.~\refcite{1166}); in particular, mutually exclusive classical possibilities (e.g.\ in the real world) are
negations of each other and $\neg \neg P$ is equivalent to $P$, just as in classical
logic. But unlike classical logic, $P \vee \neg P$ is \emph{not} a theorem of QM, and
unlike intuitionistic logic, $\neg (P \& \neg P)$ is also \emph{not} a theorem of QM;
as noted in Sec.~\ref{mp}, there must exist a nonclassical model $\mathcal{M}$ for QM
in which $P \& \neg P$ is the case. The interpretation of $P \& \neg P$ in $\mathcal{M}$
is identical to that explained in Secs.~\ref{qs} and~\ref{sch}; `$P$'~(resp.~`$\neg P$') of
$P \& \neg P$ has the nonclassical meaning that `$\neg P$'~(resp.~`$P$') is not provable
in the observer's interpretation QM* of QM. In keeping with the informal convention noted in
Secs.~\ref{qs} and~\ref{sch}, the observer agrees to keep his axiomatic assertions in tune with his
measurements in the real world, so that the superposition $P \& \neg P$ in $\mathcal{M}$ also has
an equivalent nonclassical meaning with `$P$'~(resp.~`$\neg P$') now denoting that ``The observer
has not measured the photon to pass through slit~L~(resp.~slit~U) at location~$\sigma$ and
time~$t_0$ in Fig.~\ref{f1}.'' One can see that the NAFL interpretation of $P \& \neg P$ is
meaningful in the real world because $\neg P$ is not \emph{really} the negation of $P$ in $\mathcal{M}$;
if the observer has not measured (axiomatically declared) that the photon passed through slit~L, it does not
follow that he has measured (axiomatically declared) that the photon passed through slit~U.
Consistency demands that the observer should never be able to \emph{prove}
$P \& \neg P$ in any NAFL theory, in particular, QM or QM*; for such a proof would imply
the contradiction: ``The photon \emph{really} passed through both slits''. The only reality
that exists, as far as the observer is concerned, is that he has not measured
or axiomatically asserted either $P$ or $\neg P$ in the real world. This same reality is
accurately modeled in the nonclassical model $\mathcal{M}$ where, somewhat paradoxically,
$P$, $\neg P$ and $P \& \neg P$ all hold, but with the nonclassical interpretations noted
above; the existence of such a nonclassical model for QM is a requirement of consistency
in NAFL.

NAFL requires that both $P \& \neg P$ and its negation $\neg (P \& \neg P)$
(or equivalently,  $P \vee \neg P$) are not legitimate propositions in the theory syntax
of QM; see Sec.~\ref{tsps}. What this means is that \emph{formally}, the status of
the photon in the theory syntax of QM is indeterminate; it cannot be proven to be
either a particle (in which case $P \nolinebreak \vee \nolinebreak \neg P$ should be
a theorem) or a wave (in which case $P \& \neg P$ should be a theorem). Consequently, both
classical and nonclassical models, with respect to the proposition~$P$, exist for QM.
Note that the NAFL semantics for the nonclassical model $\mathcal{M}$ of QM does not
take a stand either on the status of the photon as a particle or non-particle.
Nevertheless, the NAFL theory QM tacitly supports (but not requires) the `reality' of
the particle nature of the photon in a \emph{metalogical} sense, \emph{i.e.}, outside of
both theory syntax and semantics. This will be fully explained in Sec.~\ref{afsh}. Here we
observe that the proof syntax of QM requires the deduction of $P \vee \neg P$;
see Sec.~\ref{tsps}, where it is noted that the rules of inference of NAFL theories,
which determine the proof syntax, must be classical. It is possible (but not necessary)
to interpret this deduction as a metalogical assertion of the particle nature of
the photon in the sense that the photon `really' had to pass through one and only
one of the two slits U and L at time~$t_0$ in Fig.~\ref{f1}, even though, at that
instant, the observer did not have a proof that either of these paths was traversed.
It is this lack of knowledge at time~$t_0$ that NAFL semantics expresses as a requirement
of logical consistency, via the nonclassical model $\mathcal{M}$, rather than any
perceived reality for the particle nature of the photon. But such a perceived reality became inevitable
the moment Definition~\ref{defnegp} was formulated in QM as the negation of $P$. Here it should be kept in
mind that the `photon' referred to is that which must necessarily pass through at least one of the slits.
The coin toss experiment, considered in Sec.~\ref{ctoss}, will further illustrate the metalogical `reality'
of $P \vee \neg P$ in NAFL.

\subsection{Bohr Complementarity in Afshar's experiment}\label{afsh}
Consider again the single-photon version of Afshar's experiment in Fig.~\ref{f1}, with $P$,
$\neg P$ and the theory QM as defined in Sec.~\ref{qsae}. At time $t = t_0$ a photon passes
through the slit(s); at $t = t_1 > t_0$, the said photon passes the wire grid in the
``interference plane'' at location~$\sigma_1$, and subsequently passes through the
converging lens; at $t = t_2 > t_1$, the photon ends up at one of the locations marked
$\mbox{U}^{\prime}$ or $\mbox{L}^{\prime}$ (say, $\mbox{U}^{\prime}$, for the sake of
definiteness). Note that in the single-photon version of Afshar's experiment, the
restriction is that only one photon at a time can pass through the slit(s), although
several photons actually end up at $\mbox{U}^{\prime}$ and $\mbox{L}^{\prime}$.
\begin{definition}\label{defq}
Take the proposition $Q$ to denote that ``The photon reaches the image $\mbox{U}^{\prime}$ of
slit U at location $\sigma_2$ and time $t_2$ in Fig.~\ref{f1}''.
\end{definition}
\begin{definition}\label{defr}
Let the proposition $R$ denote that ``The images $\mbox{L}^{\prime}$ and $\mbox{U}^{\prime}$
of the slits L and U respectively are undistorted and unchanged in intensity when the
wire grid is inserted at the calculated minima of the expected interference pattern at
location~$\sigma_1$ in Fig.~\ref{f1}''.
\end{definition}
Consider the following contentious inferences that are inherent in the controversy
over Afshar's experiment.
\begin{inference}\label{inf1}
The theory QM+$Q$ proves $P$.
\end{inference}
This, of course, implies the particle nature of the photon. 
\begin{inference}\label{inf2}
The theory QM+$R$ proves $P \& \neg P$.
\end{inference}
In words, Inference~\ref{inf2} means ``The photon passed through both slits U and L at location~$\sigma$ and
time~$t_0$ in Fig.~\ref{f1}'', which in turn implies the wave nature of the photon. Since $Q$
and $R$ are true, in the sense that they are observed, one may conclude that if
Inferences~\ref{inf1} and~\ref{inf2} are granted, the resulting `true' (but inconsistent)
theory QM+$Q$+$R$ violates BCP; see Definition~\ref{bcp}. Kastner~\cite{kastner1,kastner2} seems to
permit Inference~\ref{inf1}, but disputes that it `really' establishes `which-way' information
for the photon; she argues that Cramer's Transactional Interpretation~\cite{cramer} supports
her stand. Kastner concludes that only the wave nature of the photon is unambiguously
exhibited at the slits in Afshar's experiment, via Inference~\ref{inf2}, which she allows
(although neither she nor Afshar interpret the wave nature of the photon as a proposition of
the form $P \& \neg P$, as NAFL requires). In what follows, we will argue that the NAFL
interpretation allows Inference~\ref{inf1}, but \emph{not} Inference~\ref{inf2};
this fact, when coupled with the temporal nature of mathematical truth in NAFL, means that
Bohr complementarity survives. The claim that a single photon `really' exhibits wave nature
is questionable from the NAFL point of view and will be critically examined in
Sec.~\ref{noinf2}. However, we emphasize at the outset that the wave nature of light
could well follow unambiguously in a new, as yet unknown, version of QM formalized in NAFL.
The present argument applies only to our current understanding of QM and the nature of light.

The NAFL justification of BCP is as follows. For $t_0 \le t < \nolinebreak t_2$, the observer
has the theory QM in mind with the interpretation QM*=QM. Hence by Proposition~\ref{p1} (see
Sec.~\ref{mp}), $P \& \neg P$ holds for the observer, in a nonclassical model $\mathcal{M}$
for QM. At $t=t_2$, upon measuring $Q$, the observer switches to the interpretation \linebreak
QM*=QM+$Q$ and concludes $P$ in a classical model for QM, via a proof in QM*
(Inference~\ref{inf1}). Note that $P$ only applies retroactively, for times $t \ge t_2$;
therefore it does not temporally conflict with the superposition $P \& \neg P$, which
applied for $t_0 \le t < t_2$.  The second observation here is that the theory QM does not
\emph{prove} $P \& \neg P$; as noted earlier, $P \& \neg P$ is not even a legitimate
proposition in the theory syntax of QM. Such a proof of $P \& \neg P$ would make QM
inconsistent in NAFL, which is \emph{not} a conventional paraconsistent logic, as noted in Sec.~\ref{mp}.
In fact, the Main Postulate of NAFL and consequently, the existence of $\mathcal{M}$, are
strictly metamathematical results, \emph{i.e.}, pertaining to semantics
or model theory; these concepts are not formalizable in either the theory syntax or proof
syntax of QM. The theory QM* can therefore prove $P$ without loss of
consistency. Since the deduction of $P$ and the meta-deduction of the superposition
$P \& \neg P$ were made in non-overlapping time intervals, and, in particular,
were \emph{not} made \emph{simultaneously}, BCP survives in NAFL.

One might object that there is something unsatisfactory about this
state of affairs; if we interpret $P$ and $P \& \neg P$ as affirming the particle and the
wave natures of the photon respectively, they seem to be contradictory even if asserted at
non-overlapping time intervals, since, after all, they both apply to the same photon at the
same spatial location and time; this is in fact the essence of Afshar's argument~\cite{afshar1}.
The answer to this objection is that in the nonclassical
model $\mathcal{M}$, $P \& \neg P$ only means that the observer has not axiomatically
asserted $P$ via a proof in QM* (or measured $P$ in the real world) and he has not
axiomatically asserted $\neg P$ via a proof in QM* (or measured $\neg P$ in the real world);
as noted in Sec.~\ref{qsae}, $P \& \neg P$ does \emph{not} imply that the photon `really'
exhibited wave nature and passed through both slits. But the retroactive assertion of
$P$ (via a proof in QM*), valid for $t \ge t_2$, can be taken to have the \emph{metalogical}
(see Sec.~\ref{qsae}) meaning that the photon `really' passed through only
slit~U at time $t_0$, and therefore does not conflict in meaning with the meta-deduction of
$P \& \neg P$ made in $\mathcal{M}$ during the time interval $t_0 \le t < t_2$; both are
indeed `true' when appropriately interpreted. To summarize, for $t_0 \le t < t_2$,
NAFL semantics only recognizes the metamathematical truth of $P \& \neg P$ in $\mathcal{M}$.
For $t \ge t_2$, NAFL semantics asserts the retroactive truth of $P$ via a classical model
for QM; this truth may be taken to hold \emph{metalogically} for $t_0 \le t < t_2$,
\emph{i.e.}, outside of NAFL syntax and semantics. The temporal nature of
NAFL truth plays a vital role in removing the mystery associated with wave-particle
duality. In classical logic, unlike NAFL, the retroactive assertion of $P$ at $t=t_2$ would
necessarily mean that $P$ was \emph{always true} in the semantics of QM, including
at $t=t_0$. Therefore in the framework of classical logic, such a retroactive assertion
of $P$ will be problematic and will clash with the quantum superposition state that
actually held in the semantics of QM at $t=t_0$.

At this stage the reader will have the following obvious question; does not
$R$~(Definition~\ref{defr}) \emph{prove} the existence of the interference pattern
at location $\sigma_1$ of Fig.~\ref{f1} and consequently, the reality of the wave
nature of the photon when it enters the slits U and L? In the NAFL interpretation, the
answer to the second part of this question is in the negative; in Sec.~\ref{noinf2} we will
argue that Inference~\ref{inf2} is not permitted on logical grounds and therefore
the observation $R$ is irrelevant to the above-noted retroactive conclusion of $P$ by
the observer. A second question that arises is whether there are logical
grounds for banning Inference~\ref{inf1} as well. Afshar~\cite{afshar1} has asserted
that Inference~\ref{inf1} follows from standard optics and has subsequently pointed out
elsewhere that moving one of the slits U or L in Fig.~\ref{f1} causes the corresponding
image $\mbox{U}^{\prime}$ or $\mbox{L}^{\prime}$ to co-move with the slit. Here we do
not pass judgement on the experimental validity of Afshar's claims. We only wish to
point out that on purely logical grounds, NAFL does permit the \emph{conclusion} of
Inference~\ref{inf1}, namely, the retroactive assertion of $P$; for reasons mentioned
in Sec.~\ref{qsae} and in this subsection, NAFL does not object to the metalogical existence
of the photon as a particle. Thus one could also take the interpretation at
$t=t_2$ as QM*=QM+$P$ (instead of QM*=QM+$Q$ as noted above), if it turns out that there is indeed
something wrong with Inference~\ref{inf1}.

We note that although the NAFL interpretation does permit the retroactive
assertion of $P$, via Inference~\ref{inf1} or otherwise, there is no \emph{obligation}
on the part of the observer to make such an assertion (or measurement). The
observer can choose to live with just the \emph{metamathematical} conclusion that both
classical and nonclassical models for QM exist; he could choose to remain agnostic
(\emph{i.e.}, take no stand) on the \emph{metalogical} status of the photon as a
particle or non-particle in the nonclassical models and scrupulously avoid making any
retroactive assertions/measurements. Consequently, to facilitate such an agnostic
attitude, Inference~\ref{inf1} must either be disallowed or if permitted, must be weakened so as not to
imply any `reality' for the path of the photon at time~$t_0$. At time~$t_2$, the photon
behaves \emph{as though} it originated from slit~U (via the weakened Inference~\ref{inf1}), but need not
have `really' done so, from the agnostic's point of view. Kastner~\cite{kastner1,kastner2}
seems to prefer this latter approach to Inference~\ref{inf1}; however, rather than remain
agnostic, she asserts that the photon passed through both slits, as a wave. In Sec.~\ref{kast},
we criticize the limiting of Inference~\ref{inf1} as noted above, as well as Kastner's
reasons for doing so. Cramer's Transactional Interpretation~\cite{cramer} is also criticized from the
NAFL point of view in Sec.~\ref{cram}. An \emph{anti-realist} approach is also possible in NAFL. Anti-realism
is stronger than agnosticism, in the sense that it requires the observer to \emph{deny} any
reality for the state of the photon as particle or non-particle in the nonclassical
model~$\mathcal{M}$. In other words, no such reality exists in the absence of an axiomatic
declaration, which, by informal convention, the observer associates with `measurement'
in the real world. The anti-realist approach would presumably require the
banning of Inference~\ref{inf1} and other retroactive assertions of reality. Of course,
Inference~\ref{inf2} is already illegal in NAFL, as will be explained in Sec.~\ref{noinf2}.

In summary, there are three approaches possible in NAFL regarding the status of the
photon during $t_0 \le t < t_2$, when $\mathcal{M}$ is in force. These are, a metalogical
reality for the particle state, agnosticism and anti-realism. The author believes
that the first option is the most satisfactory from both philosophical and logical
points of view, \emph{given} Definition~\ref{defnegp}. With such a choice of negation
in NAFL, has the observer denied any possibility that the photon can `really'
pass through both slits, thereby affirming its metalogical particle status? This is a
tricky issue; the author believes that the answer is in the positive, i.e., the observer
has predetermined the particle nature of the photon via Definition~\ref{defnegp}. One may attempt to justify
the above choice of negation by arguing that $P$ and $\neg P$ cannot be `measured' together, although
$P \& \neg P$ could be `really' true in the real world (this would provide another reason for banning
Inference~\ref{inf2}, for otherwise the proposition $R$ in Definition~\ref{defr} surely amounts to a `measurement'
of $P \& \neg P$). However, such an attempt at justification fails, for the following reason.
NAFL truth for undecidable propositions of a theory is purely axiomatic in nature, by the
Main Postulate (see Sec.~\ref{mp}); the concept of `measurement' cannot be formalized
in the theory syntax of NAFL theories~\cite{ijqi}. The informal convention of associating
`axiomatic declaration' with `measurement' in the real world can, in principle, be broken
in NAFL; see Sec.~\ref{sch} as well as the final paragraph of Sec.~\ref{noinf2}. If the photon can `really'
(or classically) pass through both slits, there ought to be nothing to stop
us from axiomatically declaring, i.e., inferring via a proof in QM*, that
`The photon passed through both slits at $t=t_0$', even if \emph{measurement} of this
event is impossible. For such a proposition to be legal in the theory syntax of QM,
NAFL would require the negation of $P$ to be modified such that it includes the case of
the photon passing through both slits in disjunction with the case noted in
Definition~\ref{defnegp}. \emph{However}, as a consequence, BCP will fail by the Main
Postulate of NAFL. For if both the particle and wave natures of the photon are classically possible
phenomena, then the formal undecidability in QM of whether the photon exists as a particle
or a wave would demand (via Proposition~\ref{p1}) that there must exist a nonclassical
model $\mathcal{N}$ for QM in which the superposed state of the photon as a particle
\emph{and} a wave must hold. In $\mathcal{N}$, the photon will be neither a particle
nor a wave, in violation of BCP. As explained in Sec.~\ref{noinf2}, we do not believe
that this is the correct approach.

Reverting to the negation in Definition~\ref{defnegp}, the ``wave nature'' of the photon
is a proposition of the form $P \& \neg P$. As noted earlier, neither $P \& \neg P$
nor its negation $P \vee \neg P$ (which symbolizes the particle nature of the photon)
is a legitimate proposition in the theory syntax of QM. Consequently, BCP survives,
because the Main Postulate and Proposition~\ref{p1} apply only to formal propositions that
are in the theory syntax of NAFL theories; the nonclassical model~$\mathcal{N}$ noted
above need not (and does not) exist. Here we have put ``wave nature'' in quotes because
$P \& \neg P$ in the nonclassical model $\mathcal{M}$ of QM does not imply that the photon
is `really' a wave, as was argued above and in Sec.~\ref{qsae}. One can also state BCP, in
the context of the NAFL interpretation of Afshar's experiment, as follows:
\begin{quote}
At any given time, the observer, via the interpretation QM*, can generate either a
classical or a nonclassical model (but \emph{not} both) for QM, with respect to the
proposition~$P$; in the classical model, $P \vee \neg P$ (and either $P$ or $\neg P$) holds,
and the photon is a particle; in the nonclassical model, $P \& \neg P$ holds and the photon
may be loosely termed as a `wave', although its true status in the real world is left
ambiguous.
\end{quote}
One can see that the above formulation of BCP is not violated in the NAFL
interpretation of Afshar's experiment. The NAFL interpretation also neatly solves the
problem of the mysterious ``instantaneous collapse of the wavefunction'', which would
arise if and only if one insists that the photon `really' exhibits wave nature.
Indeed, when $Q$ is measured at time~$t_2$ in Fig.~\ref{f1}, all that happens is that
the observer switches his interpretation QM* of QM as noted previously. This amounts to a
switch from a state of ignorance regarding the path of the photon to one of knowledge.
Obviously, there is no implication here that the photon abruptly collapsed from a wave
to a particle at $t=t_2$.

\subsection{Critique of Inference~\ref{inf2}}\label{noinf2}
From the point of view of logic (and in particular, the NAFL interpretation), we wish to
establish that \emph{the presence of an interference pattern, such as, that observed in
Young's two-slit experiment, does \textbf{not prove} the `reality' of the
wave nature of a single photon}.

Let us first consider the single-photon version of Young's two-slit experiment.
Let the wire grid at location~$\sigma_1$ in Fig.~\ref{f1} be replaced by an
electronic screen, capable of registering and storing the arrival of
single photons. The photons reach the slits one at a time, with equal probability of
passing through either slit. In the standard quantum formalism, a photon is
\emph{assumed} to take all available paths to any particular spot on the screen at which
it ends up, \emph{i.e.}, the photon is assumed to pass through both slits and `interfere
with itself', in order to theoretically predict the interference fringes. Since these
predictions agree with the experimental observations, the `reality'
of the wave nature of the photon is concluded. At the outset, let us grant that the
interference fringes are indeed observed even in the single-photon case, even though
doubts have been expressed in this regard when the rate of emission of photons is
sufficiently small~(Ref.~\refcite{mardari}; see the paragraph ``If quanta are to be
treated as real particles, self-interference must be ruled out. \dots Nevertheless,
this still means that the evidence in favour of self-interference is inconclusive.'').
Firstly, note that when a single photon is fired at the slits, it also ends up as a single,
bright spot at a specific, unpredictable location on the screen; \emph{it does not exhibit
an interference pattern on the screen}. This fact conclusively establishes the grainy nature
of the photon as it interacts with the screen. The interference pattern, which is observed
to build up over time only after many photons have landed on the screen, has to be
interpreted as reflecting a probability distribution, with the probability density
function~(PDF) proportional to some power of the intensity of the fringes. In particular,
at the dark fringes of the screen, the photon has a vanishing PDF. Note that the PDF
for the photon is normalized over the entire area of the $\sigma_1$ plane in which
the screen is located.
\begin{remark}\label{r1}
A zero value of the probability density function~(PDF) for the photon, at a point
or a line located on a dark fringe (minimum) of the interference pattern on the screen
in Young's two-slit experiment, does \emph{not} imply a proof in the theory QM that a
given photon cannot land at that point or line and be detected.
\end{remark}
Remark~\ref{r1} seems to follow even in the standard formalism for QM. Note that the
probability (as opposed to the PDF) of the photon reaching \emph{any} given point or line on
the screen is \emph{exactly} zero; but obviously this does not constitute a \emph{proof}
that the photon cannot be detected at that point or line. Points/lines on which the PDF
vanishes, such as, the dark fringes in the two-slit experiment, are not special in this
regard. An \emph{arbitrarily} small area around such a point or line will still have a
non-zero probability of recording a photon, as long as the PDF does not vanish identically
in the entire area. So there does not appear to be any basis for excluding the possibility
that a photon can land at the minima of the fringes, although one could expect that such an
eventuality is unlikely in the real world. But one should not confuse
`statistical expectation' with `proof'. To be sure, the observed intensity at the
dark fringes is zero, and zero intensity means zero photon count rate.
But remember that we are asking if a \emph{single, given} photon can land at the dark
fringes; `count rate' and `intensity' and even `probability' are not really well-defined
for this process. For example, a very large number of photons~($N$) could be fired at
the slits and $N-1$ of these could conform precisely to the expected probability
distribution; if the remaining photon ends up at a dark fringe, that does not constitute
a violation of any law of QM. For as $N$ is increased and no further deviations are
recorded, the probability distribution will asymptotically conform to the theoretical
pattern.

At this stage one might advance the argument that a zero PDF at
a point $X$ on the screen means that the photon passed through both slits and
interfered destructively with itself at $X$, making it impossible for the photon to reach
$X$. This argument requires the \emph{assumption} that the photon is `really' a wave,
from the slits all the way up to the screen, with the wavefunction `collapsing' at
the screen in order to enable the detection of the photon as a particle. But such an
assumption is at best an \emph{axiom} that is \emph{not} provable in QM.
The assertion that a photon that is fired at the slits as a particle
mysteriously transforms itself into a wave, passes through both slits, and equally
mysteriously ends up as a particle at the screen, has no credibility
when interpreted as a `reality'. In any case the axiomatic nature of this
assertion with respect to QM means that Remark~\ref{r1} still holds. For there is no
proof that the same observed probability distribution cannot be arrived at by other,
less mysterious routes. In summary, if one interprets the interference pattern at
the screen strictly as reflecting a probability distribution for particle-like photons,
and if one views the quantum formalism as merely one possible algorithm for deriving
it, one cannot infer any `reality' for the wave nature of the photon, as
demonstrated by Remark~\ref{r1}. Young's two-slit experiment, by itself, does
\emph{not prove} the wave nature of the photon. One can at best assert that no photon
has so far been experimentally detected at the theoretical minima of the interference
pattern. But that argument does not constitute a proof of the theoretical
impossibility of such a detection in \emph{all} future experiments, or for all
future times in a given experiment, which would be required to establish the wave
nature of the photon. Indeed, such an infinitary proposition can never be `proven'
experimentally and \emph{must} be axiomatic with respect to QM. Of course, in the
event of such an axiomatic assertion being added to QM, the wave nature
of the photon (which can then be inferred) must be formulated as a classically possible
phenomenon rather than as a contradiction of the form $P \& \neg P$; as observed in 
Sec.~\ref{afsh}, this will require a modification of Definition~\ref{defnegp}, and
consequently, NAFL will require the violation of BCP (Definition~\ref{bcp}) via a nonclassical model
$\mathcal{N}$ in which the photon is neither a particle nor a wave. We do
not believe that this is the correct approach, not only because of the mystery
associated with wavefunction collapse, but also in the light of Inference~\ref{inf1}
in Afshar's experiment. Inference~\ref{inf2} must be sacrificed, as will be argued below.

Next consider the NAFL interpretation, with Definition~\ref{defnegp} in place. Let
$X$ be a point located on a dark fringe in Young's two-slit experiment. To `prove'
that a photon leaving the slit(s) can never reach $X$, the standard QM formalism
proceeds as follows.
\begin{enumerate}
\item Assume that the photon reaches $X$ on the electronic screen. \label{a}
\item Assume $P$, assume $\neg P$; $P \& \neg P$ follows. The photon takes all available
paths to the point $X$. \label{b}
\item Associate a probability wave with the photon, in accordance with the
standard formalism. \label{c}
\item Conclude destructive interference at $X$. The photon has a zero probability density
function at $X$. \label{d}
\item Conclude from step~\ref{d} that step~\ref{a} is false, and the photon can never
reach $X$ on the electronic screen. \label{e}
\end{enumerate}
We already questioned step~\ref{e} in the preceding analysis; a vanishing PDF does not
imply proof of impossibility. Even if step~\ref{e} is granted, the above proof is
\emph{not} valid in the NAFL version of QM, whose rules of
inference are classical. For step~\ref{b} \emph{assumes} a contradiction of the
the form $P \& \neg P$. Classically, \emph{any} proposition can be inferred from a
contradiction, and a proof based on such an inference has no validity. Therefore
the conclusion (step~\ref{e}) cannot be established as a theorem of QM based on the above
`proof', either in NAFL or in classical logic. Indeed, the wave nature of the photon
is now classically impossible to prove as a theorem of QM, since it has been
formulated as a contradiction. This holds in NAFL as well; a proposition of the
form $P \& \neg P$ is not even legitimate in the theory syntax of the NAFL theory~QM
(see Sec.~\ref{tsps}), and so cannot be a theorem. Further, $P \& \neg P$ cannot
be added as an axiom to QM, say, at time~$t_1$, for the purpose of obtaining
an interpretation QM* that retroactively asserts the wave nature of the photon, in
the same manner that $Q$ was added at time~$t_2$; as noted in Sec.~\ref{tsps},
only propositions that are legitimate in the theory syntax of QM can be so added.
\emph{However}, it follows from Proposition~\ref{p1} that consistency
of QM requires the existence of a nonclassical model $\mathcal{M}$ for
QM in which $P \& \neg P$ is nonclassically true. The nonclassical interpretation of
$P \& \neg P$ in $\mathcal{M}$ was extensively discussed in Secs.~\ref{qsae}
and~\ref{afsh}. The model theory TM for $\mathcal{M}$ must be based on a paraconsistent
logic (see Sec.~\ref{mp}), for TM must prove $P \& \neg P$. In such a
paraconsistent logic, the classical result that \emph{any} proposition follows from a
contradiction does not hold. Therefore steps~\ref{a}-\ref{d} in the above proof can be validly
formulated in TM. Rather than asserting that the photon took all available paths to the dark
spot $X$ on the screen and destructively interfered with itself, we may simply conclude that
the photon did not take any path to $X$ (and indeed, is very unlikely to do so, via the
proof in TM); hence the dark spot. Thus we have a setting in which the paraconsistent
theory TM can, in principle, justify the standard quantum formalism in a nonclassical NAFL
model $\mathcal{M}$ for QM. But note that the theorems of QM, as well as those
of its interpretation QM* (which generates $\mathcal{M}$), must also be theorems
of TM; the contradictions provable in TM must involve only undecidable propositions
of QM*, such as, $P$. In particular, since infinite sets cannot exist in NAFL
theories~\cite{ract}, including QM, it follows that TM cannot permit these either.
One must develop real analysis in NAFL without infinite sets~\cite{ract}, so that one can
justify the quantum formalism in TM. One might ask, why bother with NAFL at all?
Why not just use a paraconsistent logic to begin with, and just develop
the theory TM? The problem is that in such an eventuality, we do not have any logical
principles by which we determine which are the contradictions that should be provable in TM.
For example, there would be no particular reason for BCP to hold; without
the guiding principles of NAFL, a paraconsistent logic might as well prove the
contradiction that the photon is both a wave and a particle at the same time. Thus we
would be reduced to using our arbitrary intuitions rather than the principles of logic
in determining what should and should not be provable. One may think of TM as
essentially a set of metamathematical rules that tell us how to combine various
classical models of QM so as to generate the desired nonclassical model $\mathcal{M}$
that conforms with the observer's interpretation QM* (which in turn conforms with
his observations/axiomatic declarations in the real world). Note that TM tries to
\emph{predict} the results of the `measurements' made by the observer in the real
world, via probabilistic reasoning. TM essentially tells us what QM* will probably
look like if the observer keeps his axiomatic declarations in QM* in tune with his
(statistically large number of) observations in the real world.

In summary, our thesis is that the observed interference pattern at the screen in
Young's two-slit experiment is a manifestation of a \emph{particular} nonclassical
NAFL model $\mathcal{M}$ for QM in which the count rate of the photon at each point
of the screen is in tune with the nonclassical probability distribution derived
in the theory TM of $\mathcal{M}$. There could be other nonclassical NAFL models for QM
that follow different probability distributions, but these may not be relevant to
our real world. The important achievement here is that we have conceptually justified
the use of $P \& \neg P$ in deducing the interference pattern \emph{without conceding
any physical reality} for the wave nature of the photon. As noted earlier, $P \& \neg P$
in $\mathcal{M}$ merely reflects the observer's ignorance of the path information
of the photon and is used \emph{only} for the purpose of deriving the probability
distribution; the contradictions in the paraconsistent structure $\mathcal{M}$ have no
`physical' reality. To the various bright spots on the screen, each individual photon
`really' takes only one path. The only mystery here is that this `reality' is not
revealed via a classical probability distribution when a large number of photons are fired
at the screen; instead, the resulting interference pattern revealed by Nature
seems to indirectly confirm, via a nonclassical probability distribution, the observer's
ignorance of the path information. Why does Nature act in this manner? The answer may
have something to do with the paradoxical nature of probability, which is, even
classically, a problematic notion that is dependent on the information available to
the observer (hence, `conditional' probabilities). While the photon
passes through only one slit at a time as a particle, the status of the other inactive
slit (\emph{i.e.}, whether it is open or closed) seems to impose some sort of conditionality
on the probability distribution. Perhaps formulation of real analysis in NAFL, development
of consistent postulates for QM, and then developing the paraconsistent
theory TM of $\mathcal{M}$ will shed further light on this mystery. Many of the concepts
of standard quantum theory, such as, the notion of probability, will not be formalizable
in the NAFL theory QM and will have to be dealt with in TM, and in general, at a
metamathematical level. This is understandable, since these concepts may be specific to
the real world and the model $\mathcal{M}$ which applies to it. It is also possible that
radically new concepts of space, time, and in particular, light, may be needed for a more
satisfactory description of Nature in a NAFL theory that does away with probability
altogether.

Let us now revert to Afshar's experiment. Afshar~\cite{afshar1} argues that if a classical
probability distribution had applied, the wire grid at location~$\sigma_1$ in Fig.~\ref{f1}
should have blocked about 6.6\% of the photons. Instead, the wire grid blocked fewer than
0.1\% of the photons, as expected from the probability distribution calculated using the
standard quantum formalism. Let us grant Afshar's argument and concede that the interference
pattern does indeed exist. As argued above, this does not legitimize Inference~\ref{inf2}.
In the light of the observation of $Q$ and Inference~\ref{inf1}, one concludes that the
photon still has a metalogical particle nature for $t_0 \le t < t_2$, but with a
nonclassical probability distribution corresponding to the interference pattern, as
deducible in a nonclassical model $\mathcal{M}$ for QM. In particular, there was no `destructive
interference' at the dark fringes corresponding to the locations of the wires; the photons that
ended up at the images simply missed the wires and the dark fringes.

Note that the axiomatic nature of NAFL truth allows the observer to
\emph{axiomatically declare}, via a choice of QM*=QM+$P$ at time~$t_0$ in Fig.~\ref{f1},
that a given photon passed through slit~U, even though such a \emph{measurement} was
not made at $t=t_0$. Thus the observer breaks with the informal convention of keeping
his axiomatic declarations in tune with his measurements, and generates a classical
particle model of the photon for $t_0 \le t < t_2$, instead of the nonclassical model
$\mathcal{M}$. If at $t=t_2$, the photon is found to end up at the image
$\mbox{U}^{\prime}$, this would vindicate the observer's choice made through guesswork and
free will. Thus in principle, NAFL allows the observer to bring the particle nature of the
photon within the scope of its semantics, even though in practice, continued success on
this front would require improbable guesswork on the part of the observer. Of course,
consistency of QM requires $\mathcal{M}$ to exist (see Proposition~\ref{p1}), even if
the observer does not choose it in any particular instance.

\subsection{Critique of Kastner's argument}\label{kast}
Kastner~\cite{kastner1,kastner2} has criticized the interpretation of Inference~\ref{inf1} as
retroactively asserting the particle nature of the photon. She has essentially two
reasons for her reservations, as stated below.
\begin{itemize}
\item According to Kastner, the photon exhibited wave nature at time~$t_0$ in Fig.~\ref{f1}
in the sense that it `really' passed through both slits, as confirmed by
Inference~\ref{inf2} (which she supports with the caveat that the wave
nature of the photon, being real, is not to be formulated as a contradiction
of the form $P \& \neg P$). So, as was noted in Sec.~\ref{afsh}, Kastner requires
Inference~\ref{inf1} to be limited to the assertion that the photon was post-selected
in the state of slit~U, without any retroactive implication that the photon `really'
passed through only slit~U (as a particle). In other words, Inference~\ref{inf1} does not
provide `which-way information' for the photon.
\item Kastner cites Cramer's Transactional Interpretation~(TI) of quantum
mechanics~\cite{cramer} as unambiguously showing that the photon was selected in the
superposed state of both slits at time~$t_0$, and also was post-selected
in the state of slit~U at time~$t_2$. However, there are intermediate times
in Fig.~\ref{f1}, when the photon was between the locations $\sigma_1$ and
$\sigma_2$, during which the ontological state of the photon is ambiguous
according to TI (see Fig.~3 of Ref.~\refcite{kastner1}). Kastner states that at
these locations, the `offer wave' of TI shows the photon to be in a superposition state
of both slits, while the backwards-in-time `confirmation wave' of TI shows the photon to
be in a state of slit~U. For this reason, Kastner asserts that it would be wrong to
retroactively infer at time~$t_2$ that the photon passed through only slit~U,
since such an inference would require the photon to be determinately a particle at all
intermediate locations between $\sigma$ and $\sigma_2$ in Fig.~\ref{f1}.
\end{itemize}
On the basis of the above criticisms, Kastner concludes that Afshar's experiment does
not refute BCP~(Definition~\ref{bcp}).

We have exhaustively addressed the problems with the first of Kastner's criticisms.
From the point of view of NAFL, Inference~\ref{inf1} is merely the observer's
retroactive axiomatic declaration of the particle nature of the photon. But the
concept of `axiomatic declaration' itself cannot be formalized in the theory syntax of
NAFL theories~\cite{ijqi}. So when the observer `post-selects' the photon as being in
the `state of slit~U', he can \emph{only} have in mind that the
photon `really' passed through slit~U as a particle. Therefore, in NAFL, the classical
model of QM that is generated at time~$t_2$ via the observer's interpretation QM*=QM+$Q$,
\emph{must} reflect such a retroactive implication for Inference~\ref{inf1}.
In fact, the completeness theorem of first-order logic (which is a metamathematical
principle in NAFL) \emph{requires} that there \emph{must} exist such a classical model
for QM. For, as we have already pointed out, QM does not prove either $\neg P$ or
Kastner's claim that the photon passed through both slits at time~$t_0$. This
latter claim, affirming the wave nature of the photon at time~$t_0$, is at best another
\emph{metamathematical} requirement that Kastner chooses to impose. In the conflict
between these two metamathematical requirements, the completeness theorem wins out
in NAFL; it is a sacred principle of logic that cannot be sacrificed. On the other
hand, the nonclassical NAFL model $\mathcal{M}$ of QM that affirms $P \& \neg P$ at
time~$t_0$ does not conflict with the above retroactive assertion of $P$, either
temporally or in meaning, as we have already pointed out in Secs.~\ref{qsae}~and~\ref{afsh}.
In particular, $\mathcal{M}$ only affirms the observer's lack of information on the path
of the photon at time~$t_0$, rather than Kastner's assertion that the photon `really'
passed through both slits.

The second of Kastner's objections is also problematic, in the sense that her stated
purpose of rescuing BCP is not served by her invocation of Cramer's TI. If TI requires
that the photon's ontological state is ambiguous between the locations $\sigma_1$
and $\sigma_2$ in Fig.~\ref{f1}, that amounts to a violation of BCP when the photon was
at these locations. For we have determined, using TI, that the photon was neither a
particle nor a wave when it was at these locations. So the NAFL model of QM that applies
at these times would have to be a superposition of both the particle and wave states of the
photon, \emph{i.e.}, a superposition of a classical and a nonclassical model of QM. But
such a superposition of models does not exist in NAFL and clearly violates BCP; see
Definition~\ref{bcp} and also the NAFL version of BCP as stated at the end
of Sec.~\ref{afsh}. This is not surprising, for TI requires us to ascribe
reality to the superposed state of the photon passing through both slits; this is
also Kastner's belief, as noted above. As a consequence, NAFL would \emph{require} that
Definition~\ref{defnegp} be modified to force, via Proposition~\ref{p1}, the existence
of a nonclassical model~$\mathcal{N}$ for QM that violates BCP (as was pointed out
in Sec.~\ref{afsh}).

\subsection{Critique of Cramer's TI in the delayed-choice scenario}\label{cram}
One example of a quantum mystery arises from the well-known delayed choice experiment of
Wheeler~\cite{wheeler}. In Fig.~\ref{f1}, \emph{after} a photon passes through the slit(s)
in the $\sigma$ plane at time~$t_0$, the observer could either choose to insert a screen
in the $\sigma_1$ plane, \emph{or} he could choose to allow the photon to pass through the
lens and reach the $\sigma_2$ plane. So according to Wheeler, this delayed choice
of measurement means that the observer could post-select (\emph{i.e., after} time~$t_0$)
each photon to pass through both slits, as a wave, or one slit only, as a particle.
But this amounts to choosing one's past after the event and seems paradoxical. In the
NAFL interpretation, the metalogical reality of the photons as particles
means that the delayed choice does \emph{not} influence the past; each photon always
passed through one and only one slit. The interference pattern, which one sees
on the screen after many photons are recorded on it, \emph{also} exists (but is not
measured) for the photons that end up at the $\sigma_2$ plane, as is spectacularly
confirmed in Afshar's experiment.

Cramer's Transactional Interpretation~(TI)~\cite{cramer}, on the other hand, provides
an explanation for delayed choice by positing an `atemporal' transaction that takes
place between `offer waves' and `confirmation waves', with the latter propagating
backwards in time. From the point of view of NAFL, however, which rejects the
relativistic conception of `spacetime' (see Appendix~B of Ref.~\refcite{acs}), such an atemporal
transaction is not physically possible, and neither can anything propagate backwards
in time. The temporal and axiomatic nature of NAFL truth requires absolute time, as well
as Euclidean space. Cramer's aphysical approach to delayed choice really amounts to taking
an anti-realist stand that the `past' exists if and only if, and only \emph{after},
it is clearly defined. This is confirmed by Cramer's assertion that ``No offer is a
transaction until it is a confirmed transaction'', which corresponds to Wheeler's ``No
phenomenon is a phenomenon until it is an observed phenomenon''. Such an anti-realist
stand is possible in NAFL by associating `measurement' or `observation' with `axiomatic
declaration', and denying any metalogical reality outside of NAFL syntax and semantics
(as was noted in Sec.~\ref{afsh}). But in NAFL, the superposed state is not really a
physical state of the photon passing through both slits, as is assumed by Cramer; the
anti-realist stand, when imposed upon the NAFL interpretation, would further require that
the photon has no determinate state in the real world when it is deemed to be in a quantum
superposition of passing through both slits. Hence from the point of view of NAFL, Cramer's
TI is not consistent with either realism or anti-realism. As was discussed in Sec.~\ref{afsh},
positing a metalogical reality for the particle nature of the photon is not only
compatible with NAFL, but is also philosophically a
more satisfactory resolution of the paradoxes associated with Afshar's experiment. From this
realist point of view, Cramer's TI, by positing waves propagating backwards in time, seems
to accept that one \emph{can} influence the past; this is denied in the NAFL interpretation
as aphysical. In Sec.~\ref{kast}, we have criticized Kastner's~\cite{kastner1} defence of
BCP, using Cramer's TI, as logically problematic; the said defense upholds BCP at the slits
by requiring its violation elsewhere, at least according to the NAFL interpretation.

\section{The coin toss and Schr\"{o}dinger cat experiments}\label{ctoss}
Consider the coin toss experiment described in Sec.~2.5 of Ref.~\refcite{1166}. An observer
tosses a fair coin and, as it lands at time~$t_0$, covers the coin under the palm of
his hand without seeing the outcome. Let $P$~($\neg P$) represent ``The outcome is
`heads'~(`tails')''. Let the observer have the NAFL theory T in mind, which includes definitions
describing this coin toss experiment. Further, at $t=t_2$, the observer lifts his hand and
sees the outcome, say, `heads'. For $t_0 \le t < t_2$, the observer chooses the interpretation
T*=T. Hence $P \& \neg P$ holds for the observer in a nonclassical model $\mathcal{T}$ for
T, signifying that he has not measured~(axiomatically declared) the outcome to be
either `heads' or `tails' during this time interval. Let $R$ denote the proposition that
``The outcome is `heads' for $t \ge t_2$''. For $t \ge t_2$, the observer takes T*=T+$R$,
in tune with his observation, so that $P$ (which is provable in T* for these times) holds
in a classical model for T. Let the proposition $Q$, formulated for $t \ge t_2$, denote
``The outcome was `heads' during $t_0 \le t < t_2$''. Our contention is
that the theory T*=T+$R$ proves $Q$, so that the observer has retroactively
asserted at $t=t_2$ that the outcome was always `heads' during $t_0 \le t < t_2$.

Is the inference of $Q$ in T+$R$ legal? In this case, the observer can feel confident that
the outcome `heads' must be metalogically~(`really') true for $t_0 \le t < t_2$.
The observer \emph{knew} that the coin was flat under the palm of his hands during
$t_0 \le t < t_2$, but as was noted in Sec.~2.5 of Ref.~\refcite{1166}, this fact, being
provably equivalent to $P \vee \neg P$ in the proof syntax of T, \emph{cannot} be
formalized as a legal proposition in the theory syntax of T. Such an (illegal)
formalization would force $P \vee \neg P$ to be a theorem of T, and prevent the
existence of the nonclassical model $\mathcal{T}$ required by Proposition~\ref{p1}
for consistency of T in NAFL. The reason this example is interesting is that
the observer \emph{knows} that the superposed state of `heads and tails' in $\mathcal{T}$
has no `physical' reality, but nevertheless it correctly reflects the observer's
ignorance of the outcome during $t_0 \le t < t_2$. In Afshar's experiment, the
situation is logically similar, but the observer does not have such a clear intuition
for the particle nature of the photon, and consequently, for the validity of
Inference~\ref{inf1}. But nevertheless, NAFL treats both cases similarly and in a
logically consistent manner.

Next reconsider the Schr\"odinger cat experiment described in \linebreak Sec.~\ref{sch}.
Once again the observer has the intuition that the retroactive assertion of
$U$, via an inference in the theory QM*=QM+$V$, is true in the real world. This
inference is based on the principle (deducible in the proof syntax of QM*) that if
one finds the cat to be alive at time~$t_2$ and if one knows that the cat was alive
at an earlier time~$t_0$, then the cat was alive for $t_0 < t < t_2$. Thus at
$t=t_2$, we \emph{know} that the cat was (metalogically, `really' and unambiguously)
alive during $t_0 < t < t_2$. NAFL supports such a rock-solid, unimpeachable
inference that conforms with the standard definition of `alive'.

\section{Concluding Remarks}
The NAFL interpretation upholds the Bohr complementarity principle~(BCP) in Afshar's
experiment~\cite{afshar1,afshar2,afshar3} by retroactively affirming the `real' particle status of the
photon, even while \emph{semantically} the photon was in a superposed state~($S$) in a
nonclassical model $\mathcal{M}$ of the NAFL theory~QM formalizing quantum mechanics.
However, such a `reality' for the particle state, which is outside of both the syntax and
the semantics of QM, can be said to be \emph{metalogical}. In NAFL, $S$ (or the `wave nature'
of the photon) only reflects the fact that the observer has not measured (axiomatically
asserted) the true, classical path of the photon; no \emph{physical} reality, to the effect
that the photon `really passed though both slits and interfered with itself', can be
assigned to $S$. The interference pattern in the $\sigma_1$ plane of Fig.~\ref{f1} must
be interpreted in NAFL as reflecting a nonclassical probability distribution for the photons,
still treated as particles, that is derivable within $\mathcal{M}$. We have no explanation
yet for \emph{why} the lack of which-way information influences the probability distribution
in this manner.

The still nascent NAFL interpretation of quantum mechanics has considerable potential
for future research. The subtle formulation of the syntax and semantics of NAFL theories to combine
both classical and intuitionistic principles, the ability of NAFL to handle the temporal nature of
mathematical truth, and the demonstration of the need for a paraconsistent logic to handle the model
theory of NAFL theories are features that make the NAFL interpretation highly suitable for the purpose of
providing a logical explanation for the mysteries of quantum mechanics, \emph{within the framework
of a single logic}. This, despite the limitations that NAFL imposes on classical infinitary
reasoning~\cite{ijqi,1166,acs,ract}, is a great advantage of The NAFL interpretation, which can be said to provide
a logical basis for many of Niels~Bohr's great physical ideas that were spelt out in the Copenhagen interpretation.
It is to be emphasized the NAFL interpretation is not \emph{essential} for the purpose of explaining the results
of the Afshar experiment and other seemingly paradoxical experiments of quantum mechanics; the Copenhagen
interpretation is adequate for this purpose. However, at present quantum mechanics cannot be
satisfactorily formalized within a single logic. For example, real analysis uses the framework of classical logic
while other quantum phenomena like superposition require nonclassical logics. If the NAFL interpretation
can be developed to its full potential, a formalization of all of quantum mechanics within a single logic
would hopefully be within reach.


\begin{thebibliography}{99}

\bibitem{afshar1} S.~S.~Afshar, E.~Flores, K.~F.~McDonald, E.~Knoesel, Paradox in wave-particle duality,
\emph{Foundations of Physics} \textbf{37} (2007) 295-305.

\bibitem{afshar2} S.~S.~Afshar, Violation of the principle of complementarity, and
its implications, in \emph{The Nature of Light: What is a Photon}, eds.~C.~Roychoudhuri and K.~Creath,
\emph{Proc. SPIE} \textbf{5866} (2005) 229-244. Preprint version available at:
http://www.irims.org/quant-ph/030503/ .

\bibitem{afshar3} S.~S.~Afshar, Violation of Bohr's complementarity: one slit or both?, \emph{AIP
Conference Proceedings} \textbf{810} (2006) 294-299. Eprint: arXiv:quant-ph/0701039.

\bibitem{greenberger} D.~M.~Greenberger, A.~Yasin, Simultaneous wave and particle knowledge in a neutron
interferometer, \emph{Phys. Lett. A} \textbf{128} (1988) 391-394.

\bibitem{englert} B.~G.~Englert, Fringe visibility and which-way information: an inequality,
\emph{Phys. Rev. Lett.} \textbf{77} (1996) 2154-2157.

\bibitem{kastner1} R.~E.~Kastner, Why the Afshar experiment does not refute
complementarity, \emph{Studies In History and Philosophy of Science Part B: Studies In History and
Philosophy of Modern Physics} \textbf{36} (2005) 649-658. Eprint:
arXiv:quant-ph/0502021.

\bibitem{kastner2} R.~E.~Kastner, On visibility in the Afshar two-slit experiment, \emph{Foundations of Physics} \textbf{39} (2009) 1139-1144. Eprint: arXiv:0801.4757 [quant-ph].

\bibitem{qureshi} T.~Qureshi, Complementarity and the Afshar experiment, arXiv:quant-ph/0701109.

\bibitem{reitzner} D.~Reitzner, Comment on Afshar's experiments, arXiv:quant-ph/0701152.

\bibitem{drezet} A.~Drezet, Complementarity and Afshar's experiment, arXiv:quant-ph/0508091.

\bibitem{steuernagel} O.~Steuernagel, Afshar's experiment does not show a violation of complementarity,
arXiv:quant-ph/0512123.

\bibitem{srikanth} R.~Srikanth, Physical reality and the complementarity principle, arXiv:quant-ph/0102009.

\bibitem{flores} E.~Flores, E.~Knoesel, Why Kastner analysis does not apply to a modified Afshar experiment,
arXiv:quant-ph/0702210.

\bibitem{ohara} P.~O'Hara, Entanglement and quantum interference, arXiv:quant-ph/0608202.

\bibitem{ijqi} R.~Srinivasan, The quantum superposition principle justified in
a new non-Aristotelian finitary logic, \emph{International Journal of Quantum
Information} \textbf{3} (2005) 263-267; \emph{Proc. Foundations
of Quantum Information}, University of Camerino, Italy,
April~16-19 (2004). Preprint available at: http://philsci-archive.pitt.edu/archive/00001923/ .

\bibitem{1166} R.~Srinivasan, Platonism in classical logic versus
formalism in the proposed~non-Aristotelian finitary logic, Philosophy of
Science Archive, Preprint ID Code~1166. Available at:
http://philsci-archive.pitt.edu/archive/00001166/ .

\bibitem{acs} R.~Srinivasan and H.~P.~Raghunandan, On the existence of truly
autonomic computing systems and the link with quantum computing, arXiv:cs.LO/0411094.

\bibitem{ract} R.~Srinivasan and H.~P.~Raghunandan, Foundations of real analysis and
computability theory in non-Aristotelian finitary logic, arXiv:math.LO/0506475.

\bibitem{cramer} J.~G~Cramer, The Transactional Interpretation of Quantum
Mechanics, \emph{Rev. Mod. Phys.} \textbf{58} (1986) 647-688.

\bibitem{wheeler} J. A. Wheeler, The ``past'' and the ``delayed-choice'' double-slit
experiment, in \emph{Mathematical Foundations of Quantum Theory}, ed.~A.~R.~Marlow
(Academic Press, New York, 1978), pp.~9-48. Reprinted in \emph{Quantum Theory
and Measurement}, eds.~J.~A.~Wheeler and W.~H.~Zurek (Princeton University Press,
New Jersey, 1983), pp.~182-213.

\bibitem{scully} M.~O.~Scully, Y.~H.~Kim, R.~Yu, S.~P.~Kulik and Y.~H.~Shih, A delayed-choice quantum eraser,
\emph{Phys. Rev. Lett.} \textbf{84} (2000) 1-5.

\bibitem{mardari} G.~N.~Mardari, An alternative to quantum complementarity,
arXiv:quant-ph/0409197.

\end{thebibliography}
\end{document}